\documentclass[reprint,superscriptaddress,amsmath,amssymb,aps,prl]{revtex4-2}
\usepackage{lineno}\usepackage[normalem]{ulem}
\usepackage{xcolor}\usepackage{tikz}
\usepackage{amsmath,amssymb}
\usepackage{MnSymbol}
\usepackage{graphicx}
\usepackage{dcolumn}
\usepackage{bm}
\usepackage[sort&compress]{natbib}
\usepackage[export]{adjustbox}
\usepackage{xcolor}\usepackage{tikz}
\usepackage{amsmath,amssymb}
\usepackage{MnSymbol}
\usepackage{graphicx}
\usepackage{dcolumn}
\usepackage{bm}
\usepackage[sort&compress]{natbib}
\usepackage[export]{adjustbox}
\usepackage{ragged2e}
\usepackage{etoolbox}

\usepackage{hyperref}

\newcommand{\eps}{\epsilon}
\newcommand{\bra}[1]{\left(#1\right)}\newcommand{\Bra}[1]{\left[#1\right]}
\newcommand{\BRA}[1]{\left\{#1\right\}}\newcommand{\ABS}[1]{\left |#1\right |}

\newcommand{\vP}{{\bf P}}
\newcommand{\vp}{{\bf p}}
\newcommand{\mL}[2]{\mathcal{L}_{#1#2}}
  
\newcommand{\dt}{\text{d}{t}}

\newcommand{\dparf}[2]{\frac{\partial^2 #1}{\partial #2^2}}

\begin{document}

\title{Nonuniform relaxation oscillations near a saddle-node-infinite-period bifurcation}

\author{Edgar Knobloch}\email{knobloch@berkeley.edu}
\affiliation{Department of Physics, University of California, Berkeley, California 94720, USA}

\author{Arik Yochelis}\email{yochelis@bgu.ac.il}
\affiliation{Swiss Institute for Dryland Environmental and Energy Research, Blaustein Institutes for Desert Research, Ben-Gurion University of the Negev, Sede Boqer Campus, Midreshet Ben-Gurion 8499000, Israel}%
\affiliation{Department of Physics, Ben-Gurion University of the Negev, Be'er Sheva 8410501, Israel}%

\date{\today}

\begin{abstract} 

Properties of spatially dependent relaxation oscillations near a saddle-node-infinite-period (SNIPER) bifurcation are described. A SNIPER bifurcation creates a large-amplitude long-period periodic orbit via the annihilation of a pair of fixed points in a saddle-node bifurcation. We show that in non-gradient spatially extended media, this orbit may undergo a long-wavelength instability {in the oscillatory regime and a finite-wavenumber (Turing-like) instability after passing the saddle-node bifurcation}, leading to spatially modulated oscillations that persist on both sides of the SNIPER. The oscillations take different forms depending on the system: a chimera state in a theta-reaction-diffusion model, and chaotic spiking in an activator-inhibitor-substrate model. The results are expected to have applications in a number of physical systems exhibiting SNIPER bifurcations, ranging from models of the nervous system through chemical reactions to nonlinear optics.

\end{abstract}

\maketitle

Nonlinear oscillations, whether spontaneous or driven~\cite{von1995principal,jenkins2013self,tyson2022belousov}, are described in terms of limit cycles that arise through bifurcations that are either local or global. Such oscillations arise in non-gradient systems under far-from-equilibrium conditions~\cite{ch93,walgraef2012spatio,meron2015book,pismen2023patterns} that preclude relaxation to an energy minimum. For small amplitude oscillations, stability criteria can be derived from weakly nonlinear theory~\cite{aranson2002world,garcia2012complex}, while Floquet theory is required in the strongly nonlinear regime~\cite{hartman2002ordinary,ali1999local,klausmeier2008floquet,kunzostov2023elements}. This is the case for both nonlinear ordinary differential equations (ODEs) and nonlinear partial differential equations (PDEs), although in the latter case, stability with respect to space-dependent perturbations must also be considered~\cite{knobloch1986oscillatory,knobloch1992nonlocal,risler2001direct,aranson2002world,garcia2012complex}. In particular, stability determination for spatially uniform oscillations emerging from global bifurcations in PDEs poses a significant computational challenge~\cite{kuramoto1984chemical,tuckerman2020computational} since it typically requires numerical evaluation of the space- and time-dependent \textit{monodromy} matrix~\cite{hartman2002ordinary,fairgrieve1991ok,pena2007two,juma2026using} or equivalently the solution of the fundamental matrix adjoint problem~\cite{gallay2011diffusive}. Consequently, stability properties of large amplitude, spatially extended oscillations remain less understood, despite their importance in applications that range from intracellular waves~\cite{leybaert2012intercellular,beta2023actin,bement2024patterning}, through cardiac rhythmicity~\cite{irisawa1993cardiac,boullin2005development}, to cavity solitons in nonlinear optics~\cite{colet2008excitability,schelte2020dispersive}.

Like the subcritical Hopf bifurcation, the saddle-node-infinite-period (SNIPER) bifurcation generates large amplitude, strongly nonlinear oscillations but these are of relaxation type, with a slow and a fast phase, and of long period~\cite{strogatz2018nonlinear}. The SNIPER bifurcation is characteristic of excitable type-I systems
~\cite{hodgkin1948local,noszticzius1985measurement,bar1987relevance,rinzel1989analysis,yeung1998nonlinear,laing2003type,bonilla2024nonlinear,gomila2005excitability,ermentrout2019recent,prescott2022excitability,moreno2022bifurcation} and in ODE models is known to generate oscillatory and excitable behaviors on opposite sides of the SNIPER onset. Recent work has shown, however, that this dichotomy need not apply in PDE models~\cite{knobloch2024emergence}. In particular, oscillatory behavior may extend past the SNIPER bifurcation, into the parameter region where no uniform oscillations are present. 

In this Letter, we demonstrate that the onset of such nonuniform oscillations is closely linked to a long-wavelength instability of uniform oscillations preceding the SNIPER and to a subcritical finite-wavenumber instability that destabilizes the stable equilibrium near the fold. We leverage the relaxation character of the oscillations to obtain a simple yet effective condition for the loss of stability of uniform oscillations near a SNIPER and illustrate the resulting behavior using a non-gradient spatial extension of the canonical theta model in which oscillations manifest as chimera states, and Meinhardt’s multi-variable activator–inhibitor–substrate model where rogue-wave-like behavior has been reported~\cite{knobloch2024emergence}. These results provide a new perspective on the emergence of nonuniform relaxation oscillations commonly observed in spatially extended systems. 

\textit{Origin of nonuniform oscillations near a SNIPER bifurcation.--} We study reaction-diffusion (RD) systems of the form
\begin{equation}\label{eq:AI}
    \partial_{t}\,\vP=\mathcal{F}_{\mu}(\vP)+\mathcal{D}\,\partial_{xx}{\vP},
\end{equation}
where $\vP(x,t)$ is a $N$-vector of variables, $\mathcal{F}_{\mu}(\vP)$ represents reaction terms depending on a parameter $\mu$ and $\mathcal{D}$ is a constant $N\times N$ diffusion matrix. We suppose that the ODE system $\partial_{t}\,\vP=\mathcal{F}(\vP)$ undergoes a SNIPER bifurcation at $\mu=0$, i.e. that a pair of equilibria annihilates in a saddle-node or fold bifurcation as $\mu$ increases through $\mu=0$ generating a large amplitude spatially uniform periodic oscillation $\vP=\vP_{\rm osc}(t)$ in $\mu>0$, assumed stable, with period $\tau$. Thus, the emergence of nonuniform oscillations is expected to be a consequence of a \textit{diffusion-driven} instability.

To study the linear stability of $\vP_{\rm osc}(t)$ we write $\vP(x,t)=\vP_{\rm osc}(t)+\vp(x,t)$, where $\vp(x,t)$ is a spatially dependent infinitesimal perturbation obeying the linear {Fourier space} equation
\begin{equation*}
    {\dot \vp}=\mL{}{}\bra{k^2,t}\vp\equiv\Bra{\mL{0}{}\BRA{\vP_{\rm osc}(t\in[0,\tau])}-k^2\mathcal{D}}\vp\bra{k^2,t},
\end{equation*}
where $\mathcal{L}_0(\vP_{\rm osc})$ is the Jacobian of~$\mathcal{F}$.
This equation has solutions of the form $\vp(k^2,t)\propto e^{\sigma(k) t} g(k^2,t)$ for a perturbation with wavenumber $k$; both ${\cal L}_0$ and $g$ are $\tau$-periodic. The quantities $\sigma(k)$ are called Floquet exponents. When $k=0$ phase  invariance in time implies that one of these vanishes; we assume that the other $N-1$ are negative, i.e., that the $N-1$ Floquet multipliers $\lambda_j(0)\equiv e^{\sigma_j(0)\tau}$, $j=1\dots,N-1$, are inside the unit circle, $|\lambda_j|<1$, to ensure stability. In the following, we focus on the effect of $0<|k|\ll1$ on the neutral mode $j=0$. For this mode we may write $\vp(k^2,t)= \vp_0+k^2\vp_1+\cdots$, $\sigma(k)= \sigma_0+\sigma_1 k^2+\cdots$, where $\vp_0=\dot\vP_{\rm osc}$ is the solution of $\dot{\vp}_0=\mL{0}{} \vp_0$ with {Floquet multiplier $\lambda_0=1$ ($\sigma_0=0$)}. This mode is unstable to long-wavelength perturbations when \cite{kuramoto1984chemical,pena2007two,gallay2011diffusive}
\begin{equation}\label{eq:sig1}
\sigma_1\propto -\langle \vp_0,\mathcal{D}\vp_0 \rangle|_{\vP_{\rm osc}(t\in[0,\tau])}>0,   
\end{equation}
where $\langle \textbf{f},\textbf{g} \rangle\equiv\int_{\tau}\widehat{\textbf{f}} \cdot \textbf{g}\, {\text d}t$ is the inner (scalar) product, and ${\widehat \vp}_0$ is the adjoint solution to $\dot{\widehat \vp}_0=-\mL{0}{}^{\dagger}\widehat \vp_0$, satisfying the normalization ${\widehat \vp}^{\rm T}_0 \vp_0=1$. 

Near the SNIPER onset, the adjoint problem typically becomes numerically stiff~\cite{brown2004phase,ermentrout1996type}, frequently resulting in a numerically expensive evaluation of \eqref{eq:sig1}. This is so in particular when the model system contains non-polynomial terms, e.g., rational functions {(here, the Meinhardt model in Case II)}. However, owing to the slow evolution in the vicinity of the saddle-node, Eq.~\eqref{eq:sig1} can be approximated by the frozen-time equation
\begin{equation}\label{eq:sigma1_apprx}
    \widetilde \sigma_1 \propto \left. -\langle {\bf v}_0,\mathcal{D}{\bf v}_0 \rangle \right |_{\vP_{\rm osc}(t\in[0,\tau])}>0,
\end{equation}
where $\vp_0(t),\widehat{\vp}_0(t)$ are replaced by ${\bf v}_0(t),\widehat{{\bf v}}_0(t)$ solving $\mathcal{L}_0{\bf v}_0(t)= 0$, $\mathcal{L}^\dagger_0{\widehat{\bf v}}_0(t)= 0$, with the time $t$ as parameter. Although the instantaneous kernel approximation does not hold pointwise over $\tau$, Eq.~\eqref{eq:sigma1_apprx} is nonetheless accurate close to the SNIPER onset. 

The SNIPER bifurcation corresponds to a fold on the branch of spatially uniform equilibria, and hence a zero temporal eigenvalue. In extended systems, the fold corresponds to a double zero spatial eigenvalue, leading to a nearby finite-wavenumber (Turing-like) bifurcation~\cite[Fig.~16(b)]{burke2008classification}. This bifurcation is typically subcritical~\cite{burke2008classification,yochelis2008front,parra2016dark,gandhi2018,parra2022,sun2025chimera}, leading to unstable steady patterns and unstable holes in $\mu<\mu_{\rm Turing}$. The parameter interval between $\mu_{\rm Turing}$ and $\mu_{\rm SNIPER}$ thus contains no simple stable solutions, and so is a candidate for the presence of complex dynamics, cf.~\cite{higuera,knobloch2024emergence}. We show below that this is indeed the case in two systems exhibiting a SNIPER bifurcation.
\begin{figure}[tp]
	\centering
    {(a)}{\includegraphics[width=0.46\textwidth]{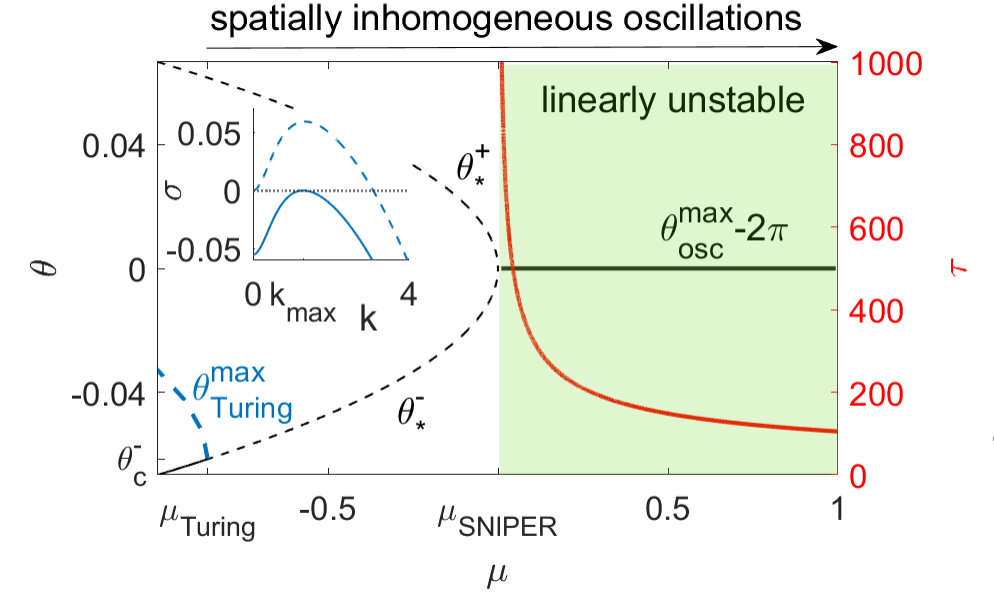}}
    {(b)}{\includegraphics[width=0.44\textwidth]{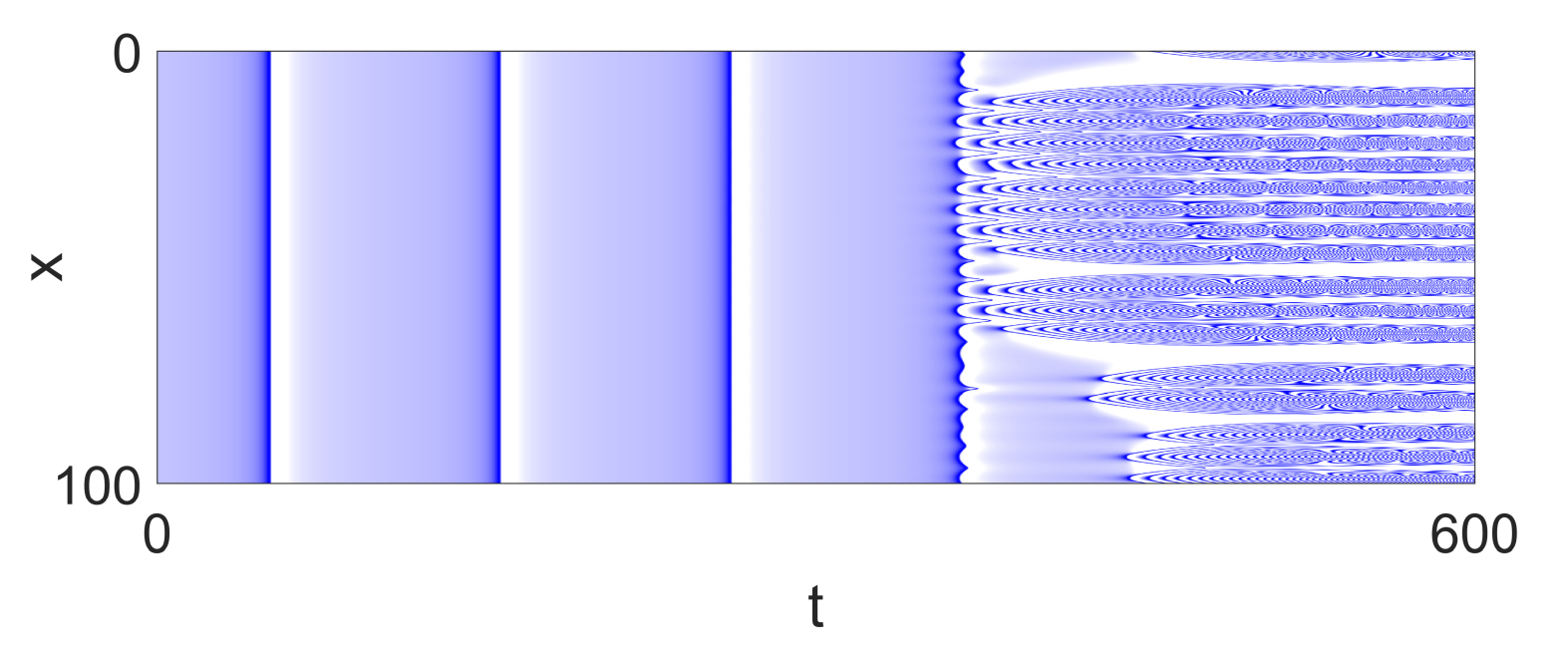}}
    {(c)}{\includegraphics[width=0.44\textwidth]{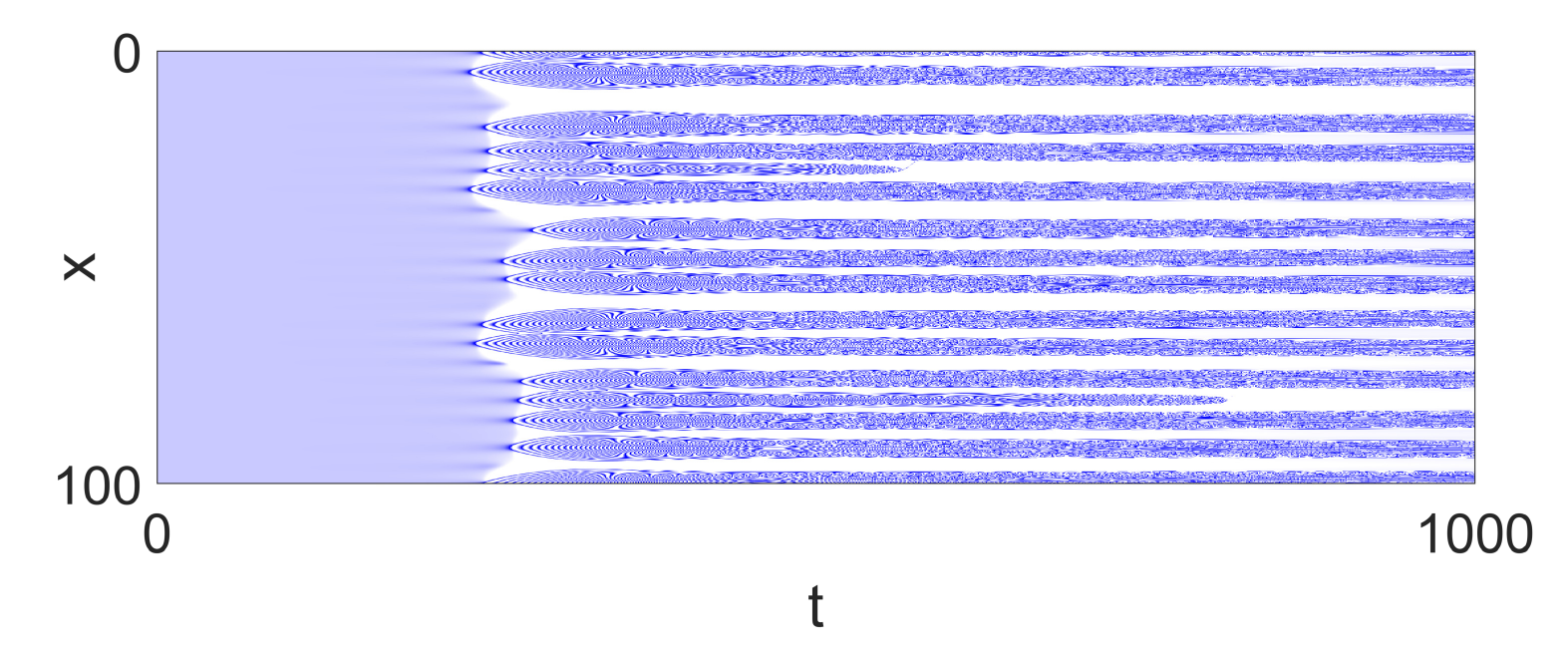}}
    {(d)}{\includegraphics[width=0.44\textwidth]{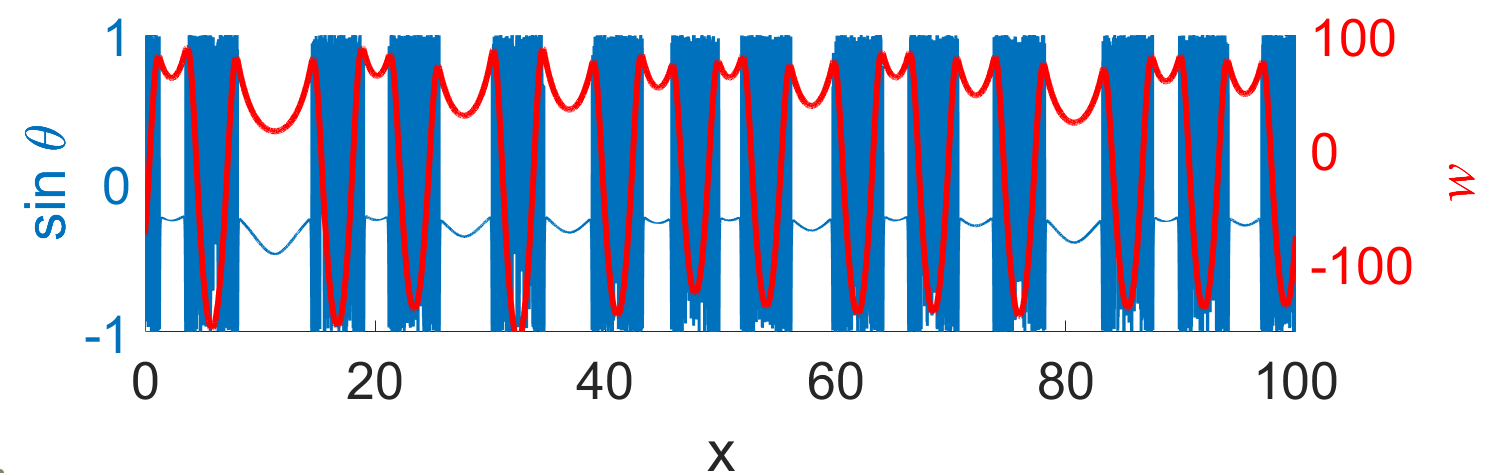}}
    \caption{(a) Bifurcation diagram showing the branches $\theta^\pm_*$ of steady uniform solutions together with the uniform oscillations $\theta_{\rm osc}(t)\in [0,2\pi)$ obtained from \eqref{eq:theta_only} using AUTO07p~\cite{doedel2012auto}, together with the steady spatially periodic Turing states ($\theta_{\rm Turing}$) that emerge from $\theta_c^-$ at $\mu_{\rm Turing}\approx -0.854$ with $k_{\rm max}\approx 1.29$; {steady spatially localized states~\cite{knobloch2015spatial} are omitted.} The
    inset shows the dispersion relations at the Turing onset (solid line) together with that at the fold (dashed line). Here $\mu_{\rm SNIPER}=0$ corresponds to the SNIPER bifurcation and solid/dashed lines indicate linearly stable/unstable equilibria. The left axis indicates the maximum values of $\theta$ while the right axis indicates the temporal period $\tau$ of $\theta_{\rm osc}$. The uniform oscillations are linearly unstable for $\mu>\mu_{\rm SNIPER}=0$ (green shaded region). {(b,c) Space-time plots of $\sin\theta(x,t)$ at (b) $\mu=10^{-3}>\mu_{\rm SNIPER}$, (c) $\mu=-10^{-4}<\mu_{\rm SNIPER}$, from direct numerical simulations~\cite{num} of~\eqref{eq:theta} with periodic boundary conditions (mod $2\pi$) and a random $\mathcal{O}(10^{-5})$ perturbation of the uniform state as initial condition. (d) Profiles of $\sin\theta(x)$ and $w(x)$ from (c) at $t=1\,000$ showing the role of $w(x)$ in the trapping of periodic oscillations. For the sake of presentation, we use $10^{3}\mu$ as a control parameter in (a), while the other parameters are $\alpha= 0.05$, $D_\theta = 0.01$, $D_{w} = 2$, $\delta = 0.1$.}}
\label{fig:theta}
\end{figure}
\begin{figure*}[tp]
	\centering
    (a){\includegraphics[width=1.\columnwidth]{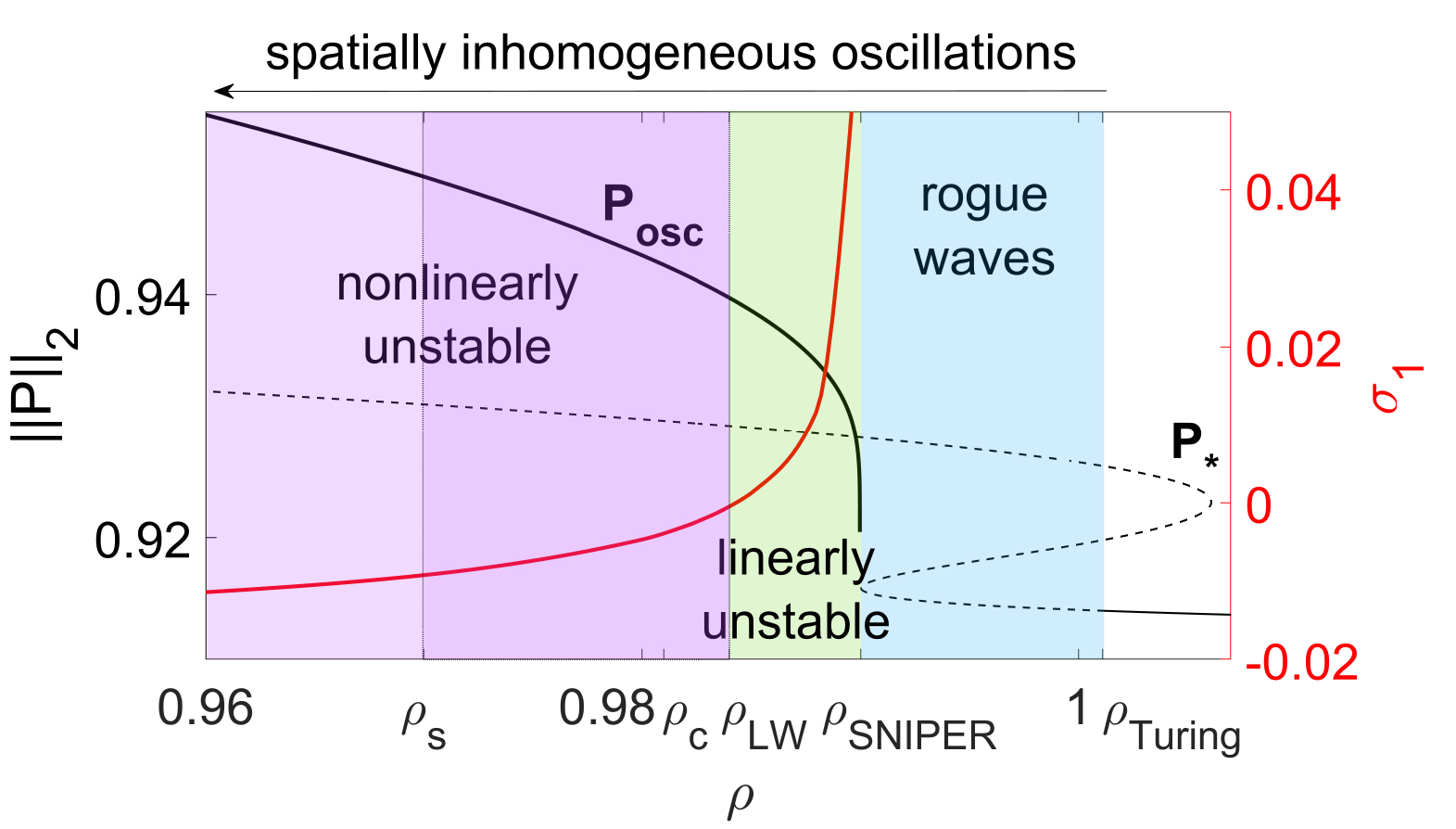}}~~
    (b){\includegraphics[width=0.95\columnwidth]{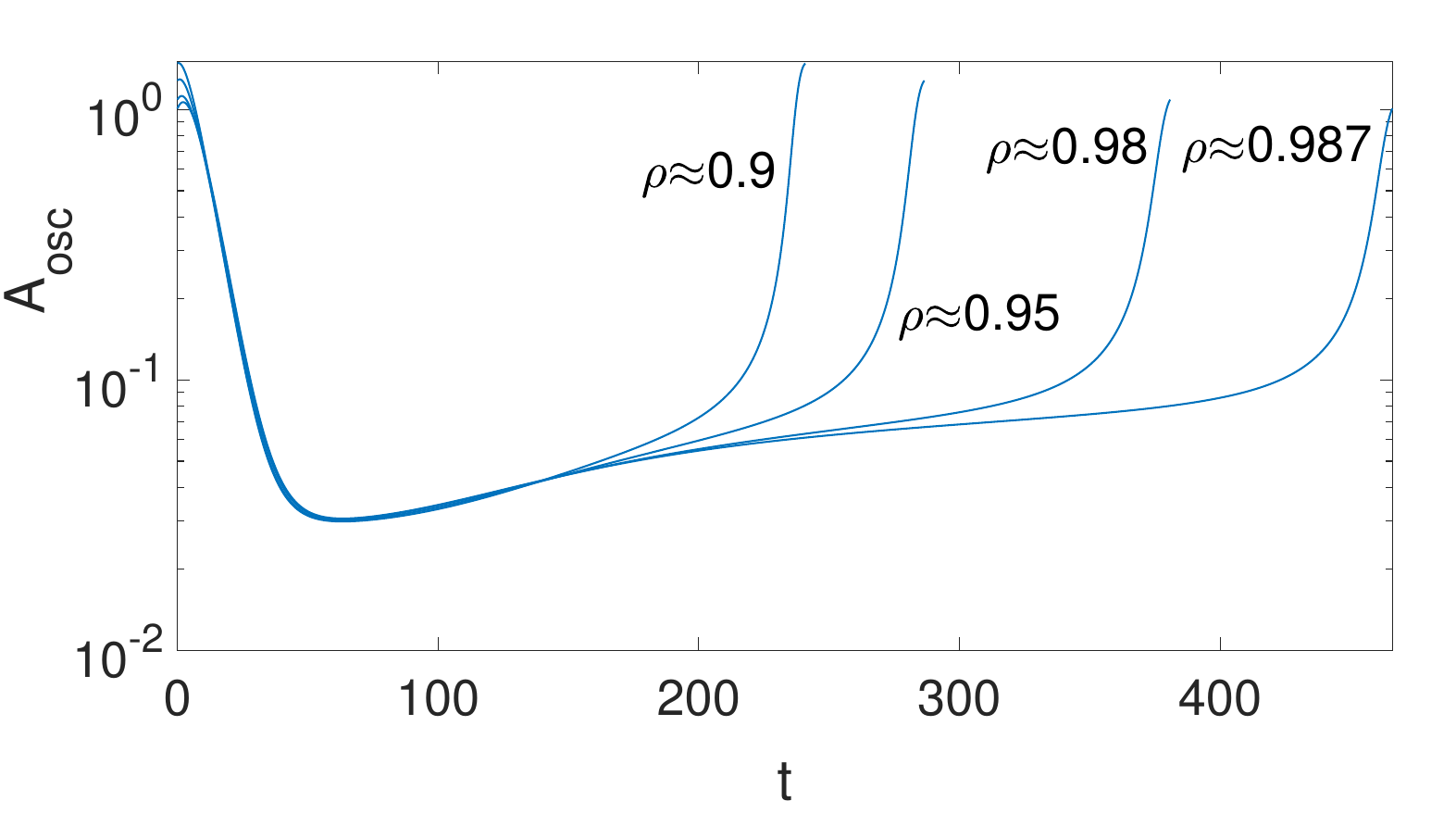}}
    \caption{(a) Bifurcation diagram for the Meinhardt model~\cite{meinhardt1976morphogenesis} showing the branch $\vP_*\equiv(A_*,H_*,S_*,Y_*)^{\rm T}$ of steady uniform solutions together with the uniform oscillatory states $\vP_{\rm osc}(t\in[0,\tau])\equiv[A_{\rm osc},H_{\rm osc},S_{\rm osc},Y_{\rm osc}]^{\rm T}$ obtained numerically using AUTO07p~\cite{doedel2012auto}, for $D_{\text{A}}=D_{\text{H}}=D_{\text{S}}=D_{\text{Y}}=0$. The SNIPER bifurcation occurs at $\rho_{\rm SNIPER}$. The left axis indicates the norm $||{\text P}||_2\equiv \sqrt{\tau^{-1}\int_\tau \dt \, {\sum_{j} P_j^2}}$, where $P_j=A,H,S,Y$ and $\tau$ is the temporal period of the limit cycle, while the right axis indicates the onset of a long-wavelength instability, $\rho\equiv\rho_{\rm LW}\approx 0.984$ and $\rho\equiv\rho_{\rm c}\approx 0.981$, obtained from Eqs.~\eqref{eq:sig1}-\eqref{eq:sigma1_apprx}, respectively. The uniform oscillations are linearly unstable in $\rho_{\rm LW}\lesssim\rho<\rho_{\rm SNIPER}\approx 0.99005$ (green shaded region) and nonlinearly unstable for $\rho<\rho_{\rm LW}$ (pink shaded region), resulting in spiking for $\rho>\rho_{\rm s}\approx 0.97$ (see text). The spiking persists into the region labeled rogue waves, $\rho_{\rm SNIPER}\approx 0.99005\lesssim\rho<\rho_{\rm Turing}\approx 1.0011$ (blue shaded region), where $\vP_{\rm osc}(t)$ does not exist. (b) Selected limit cycle profiles $\vP_{\rm osc}(t)$ showing the relaxation oscillation characteristics in terms of $A_{\rm osc}(t)$ for $\rho<\rho_{\rm SNIPER}$ plotted over one period.}
\label{fig:bif_prof}
\end{figure*}

\textit{Case {\rm I:} Chimera states in a theta-reaction-diffusion model.--} The theta model is a canonical normal form for type-I neuronal excitability capturing the onset of firing via a SNIPER bifurcation~\cite{ErmentroutKopell1986,ErmentroutTerman2008}. We combine this phase-based excitability with diffusion-driven pattern formation: 
\begin{subequations}\label{eq:theta}
\begin{align}
\label{eq:theta_only}
\partial_t\, \theta &= (1-\cos\theta) + (1+\cos\theta)(\mu-w)+D_\theta\, \partial_{xx}\, \theta, \\
\partial_t\, w &= \alpha(1-\cos\theta) - w+ D_{w} \partial_{xx}\, w+\delta\,\partial_{xx}\, \,\theta .
\label{eq:w_only}
\end{align}
\end{subequations}
Here, $\theta(x,t)$ is a phase-like variable defined mod $2\pi$ while $w(x,t)$ can be thought of as the recovery dynamics, which allows one to incorporate biologically relevant coupling mechanisms such as extracellular ionic diffusion, ephaptic interactions~\cite{wu2013role,anastassiou2015ephaptic,majhi2018alternating} (i.e., non-synaptic coupling in the nervous system), and glial-mediated feedback. The cross-diffusion term $\delta$ can be interpreted as an effective representation of gradient-driven transport of recovery variables, motivated by mechanisms observed in neural tissue and spreading depolarization phenomena~\cite{Kager2000,Dahlem2010}. 

Spatially uniform solutions of~\eqref{eq:theta} are constrained by $\cos\theta_* \in [-1,1]$ and thus correspond to
\begin{align*}
\theta^{\pm}_{*} &= {\pm \arccos \Bra{\bra{1-\mu - \sqrt{(\mu-1)^2 - 4\alpha(1+\mu-\alpha)}}/\,{2\alpha}}},\\
w^{\pm}_{*} &= \alpha(1-\cos\theta^{\pm}_{*}).
\end{align*}
For the parameters considered here, these exist for $\mu<\mu_{\rm SNIPER}=0$ with $\theta_*^-$ linearly stable to uniform perturbations [Fig.~\ref{fig:theta}(a)]. Linear stability analysis with respect to nonuniform perturbations,
\[
\begin{pmatrix}
    \theta \\
    w      
\end{pmatrix}-\begin{pmatrix}
    \theta^-_{*} \\
    w^-_{*}      
\end{pmatrix} \propto e^{\sigma t+ ikx},
\]
identifies a Turing-like instability [$\sigma(\mu_{\rm Turing},k_{\rm max})=0$, $\text{d}\sigma/\text{d}k=0$ at $k=k_{\rm max}$] at
\begin{equation}\label{eq:theta_T}
\mu_{\rm Turing} =\eta_c^2\bra{2\alpha - 1 - \eta_c^2}/\bra{1+\eta_c^2},
\end{equation}
with critical wavenumber $k_{\rm max}^2 = -G(\eta_c)/(2D_\theta D_{w})>0$, where $\eta_c\equiv\tan \theta^-_c/2$, $G(\eta)\equiv D_\theta - 2D_{w} \eta - 2\delta/(1+\eta^2)$ and $\theta^-_*=\theta^-_c$ is the solution at onset. Hence uniform solutions associated with $\theta^-_*$ are Turing unstable for $\mu_{\rm Turing}<\mu<\mu_{\rm SNIPER}=0$. Importantly, this instability is subcritical, {with steady but unstable spatially periodic and spatially localized solutions present in $\mu<\mu_{\rm Turing}$ only; neither solution type undergoes a fold,
thereby leaving a gap $\mu_{\rm Turing}<\mu<0$ in the bifurcation diagram with no simple stable solutions [Fig.~\ref{fig:theta}(a)].}
For $\mu>0$, we find uniform oscillations $(\theta_{\rm osc}(t\in[0,\tau]),w_{\rm osc}(t\in[0,\tau]))^{\rm T}$ with $\lambda_0=1$ and $\lambda_1\sim\mathcal{O}(10^{-10})$. Evaluating the linear stability of these oscillations numerically {via~\eqref{eq:sig1} shows that $\sigma_1>0$}, i.e., that uniform oscillations are stable when $k=0$ but unstable when $0<|k|\ll1$, i.e., to long-wavelength perturbations. {Figure~\ref{fig:theta} explores what happens in the gap.}

 {Figures~\ref{fig:theta}(b,c)} show space-time plots of $\sin \theta$ from direct numerical simulations (DNS)~\cite{num}. These reveal the presence, for $\mu>\mu_{\rm Turing}$, of nonuniform oscillations with intervals of periodic oscillations interspersed with intervals of chaotic oscillations resembling the so-called chimera states in spatially extended media~\cite{tobias1998convective,abrams2004chimera,omelchenko2011loss,schmidt2014coexistence,schmidt2015chimeras,laing2015chimeras,li2016spiral,clerc2016chimera,smirnov2017chimera,nicolaou2017chimera,smirnov2017chimera,laing2022chimeras,garcia2022chimera,laing2023chimeras,nicolaou2024complex}. {The latter appear to be due to the faster than exponential growth of the amplitude of the Turing state (wavelength $2\pi/k_{\rm max}\approx 4.7$) when $\mu>\mu_{\rm Turing}$}. When the amplitude of this growing state exceeds locally a threshold value, the state breaks up into incoherent or chaotic oscillations, with no coarsening detected. This process is observed not only in $\mu>\mu_{\rm SNIPER}=0$, where it is due to linear instability of $\theta_{\rm osc}$ [Fig.~\ref{fig:theta}(b)], but also in $\mu<\mu_{\rm SNIPER}=0$ where spatially uniform oscillations are absent [Fig.~\ref{fig:theta}(c)]. We emphasize that in the absence of cross-diffusion ($\delta=0$) and thus of the Turing instability, the oscillations do not pass the saddle node and the dynamical dichotomy known from the ODE is recovered.

We have confirmed our results through DNS of an extended system with $(X,Y,w)\equiv(\cos\theta,\sin\theta,w)$ on the hypersurface $X^2+Y^2=1$~\cite{nicolaou2024complex} over times as long as $t=10^5$. Figure~\ref{fig:theta}(d) shows a plot of $\sin \theta(x)$ at $t=1000$ and the corresponding field $w(x)$. The latter shows that in all instances the coherent oscillations are trapped between double maxima of $w(x)$. Thus it is the $w$ field that keeps the coherent and incoherent oscillations apart, and so is responsible for the presence of the chimera state. {As in related systems~\cite{wolfrum2011chimera,laing2015chimeras},
the detailed behavior in the chimera state is highly sensitive to model parameters, including the domain length, and in particular to the discretization used in the numerical implementation.}

We mention that the simpler gradient system
\begin{equation}
\theta_t=1-\cos \theta + \mu\bra{1+\cos \theta} - a \theta_{xx} - b \theta_{xxxx},
\end{equation}
where $a,b>0$, is described by a similar bifurcation diagram [see supplementary material (SM)], with spatiotemporal chaos that also straddles the SNIPER bifurcation at $\mu=0$, but in contrast to Figs.~\ref{fig:theta}(b,c) becomes homogeneous in space at large times. In this case $w\equiv\theta_{xx}$ has simple maxima (see SM) and no chimera states are present.

\textit{Case {\rm II:} Spiking in an activator-inhibitor-substrate model.--} The Meinhardt model~\cite{meinhardt1976morphogenesis}, a four-variable activator-inhibitor-substrate system that has been used to study biochemical initiation of branching~\cite{yao2007matrix,guo2014branching,guo2014mechanisms,shan2018meshwork,zhu2018turing,yochelis2021nonlinear}, exhibits related behavior. The model exhibits unexpected stochastic spiking resembling rogue waves~\cite{knobloch2024emergence} that occurs in the absence of any linear oscillatory instability. The system takes the form \eqref{eq:AI} with $\vP(x,t)\equiv[A(x,t),H(x,t),S(x,t),Y(x,t)]^{\rm T}$ representing the concentrations of an activator, an inhibitor, the substrate, and a marker for differentiation, respectively, $\mathcal{F}_{\rho}$ represents reaction terms depending on a parameter $\rho$, and $\mathcal{D}\equiv{\rm diag}[D_{\rm A},D_{\rm H},D_{\rm S},D_{\rm Y}]$ is a diagonal diffusion matrix (see SM for details).

\begin{figure}[tp]
	\centering
    {(a)}{\includegraphics[width=0.44\textwidth]{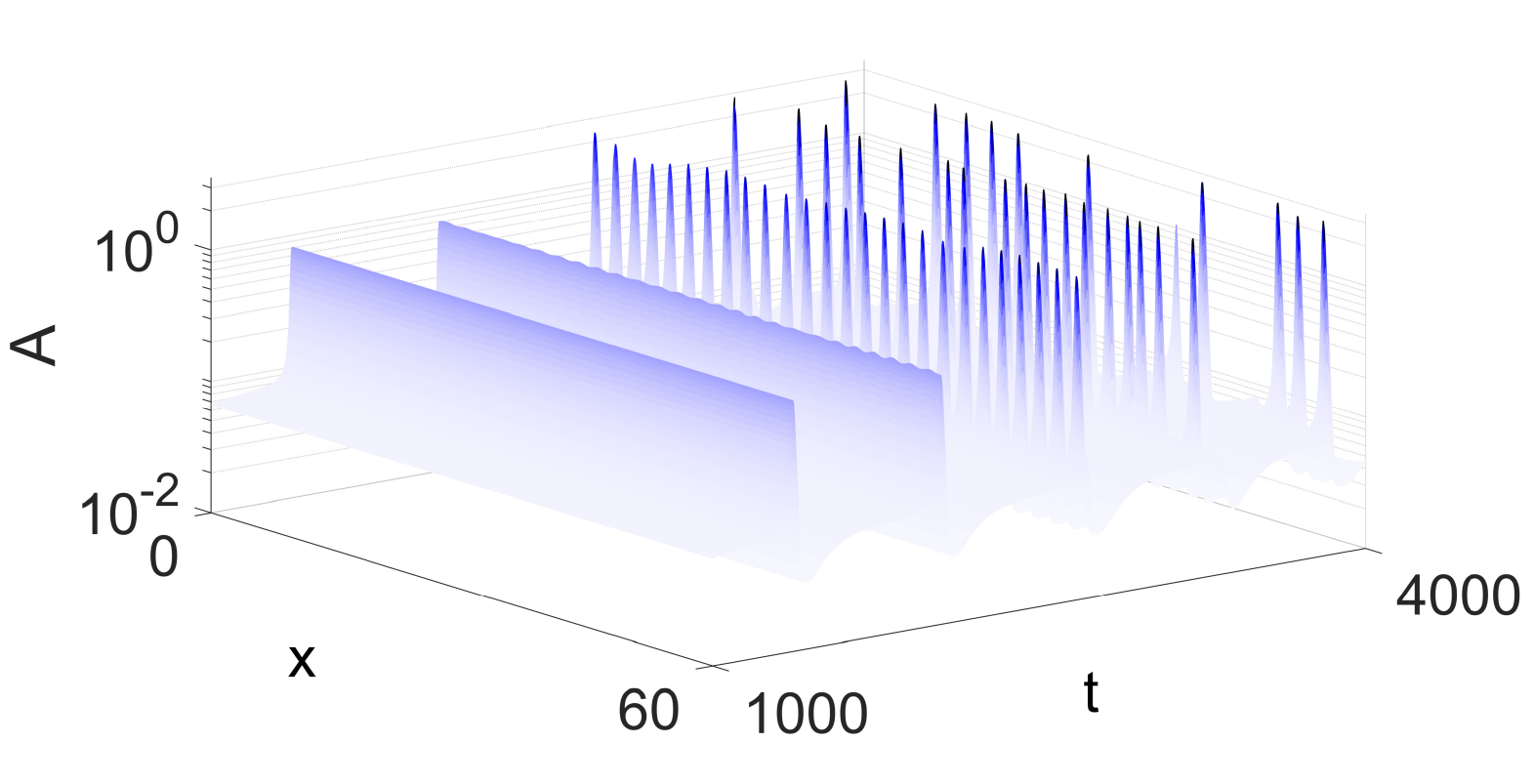}}
    {(b)}{\includegraphics[width=0.21\textwidth]{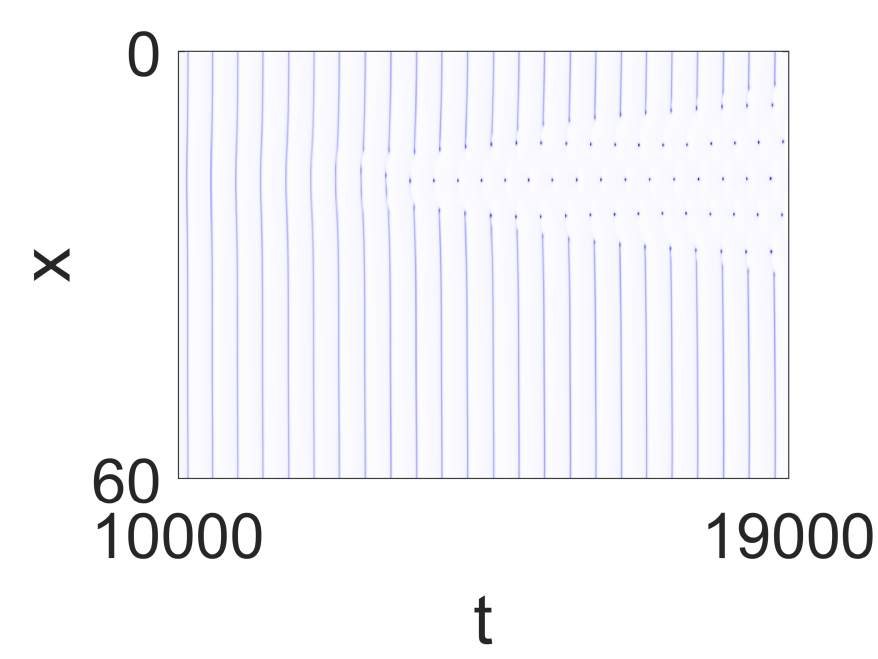}}
    {\includegraphics[width=0.21\textwidth]{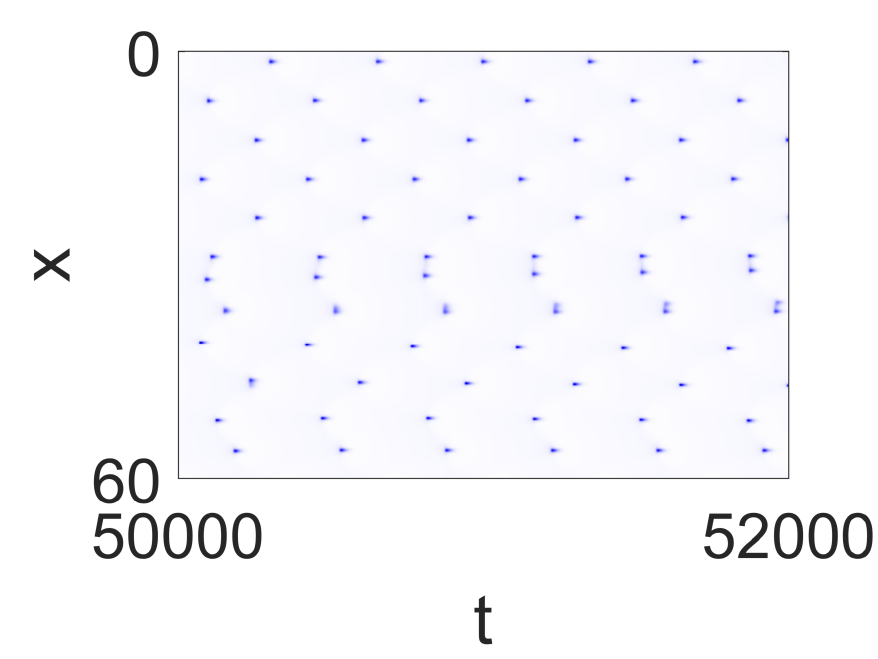}}
    {(c)}{\includegraphics[width=0.21\textwidth]{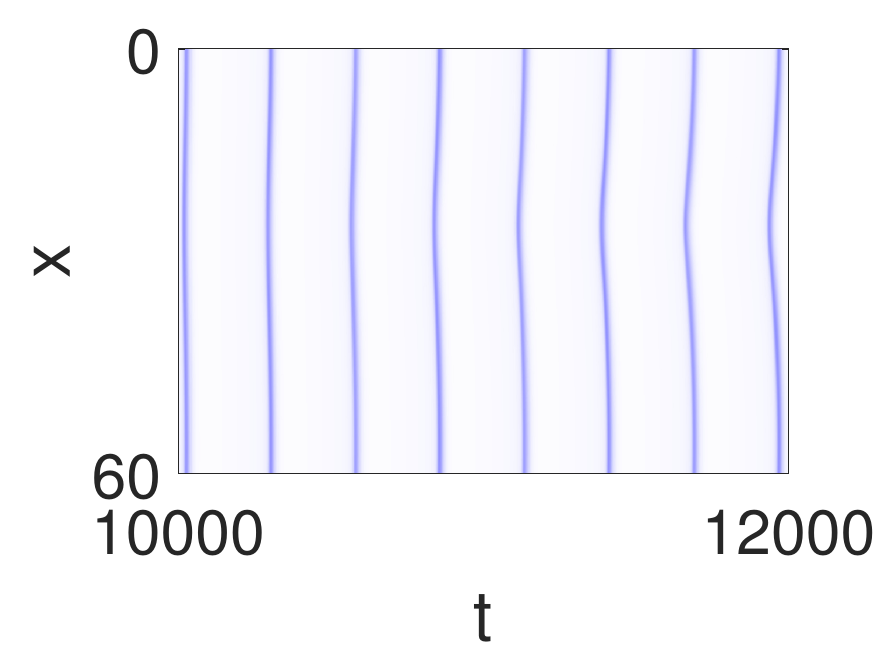}}
    {\includegraphics[width=0.21\textwidth]{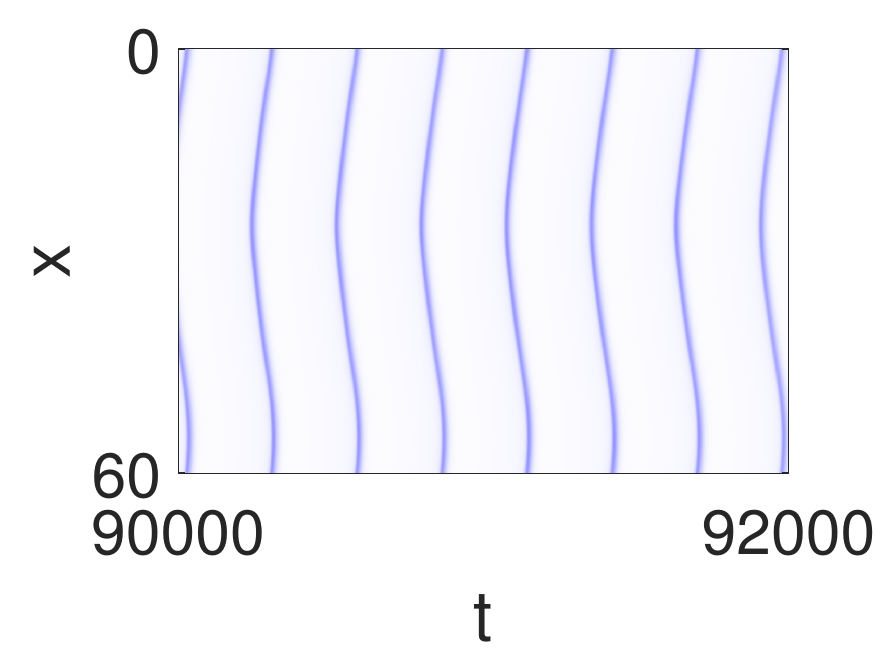}}
    {(d)}{\includegraphics[width=0.44\textwidth]{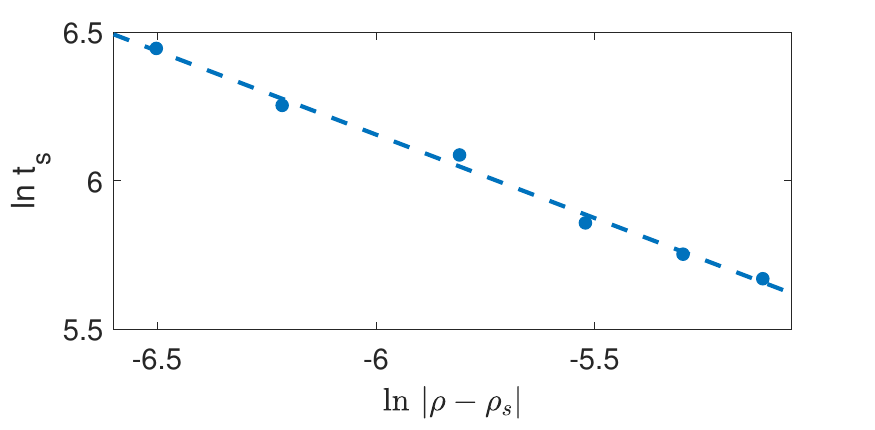}}
    \caption{
    DNS of the Meinhardt model (see SM) with periodic boundary conditions on a domain of length $L=60$, showing space-time plots of $A(x,t)$ at (a) $\rho_{\rm LW}\approx0.984<\rho=0.99 \lesssim \rho_{\rm SNIPER}\approx 0.99005$, (b) $\rho=0.98 \lesssim \rho_{\rm LW}$, and (c) $\rho=0.95<\rho_{\rm LW}$, {initialized by random perturbations of the uniform oscillation amplitude in the fields $A,S,Y$ of size (a) $\mathcal{O}(10^{-5})$, (b) $\mathcal{O}(10^{-3})$, and (c) $\mathcal{O}(10^{-1})$}; the field $H$ is not perturbed. (d) Plot of the time $t_{\rm s}$ to breakup into spiking [as in (b)] as a function of $\rho$ for $L=60$. The dashed line is a linear fit with slope $-0.56$ (and goodness-of-fit coefficient $R^2=0.99$) confirming that $t_{\rm s}\sim 1/\sqrt{|\rho-\rho_{\rm s}|}$, where $\rho_{\rm s}\approx 0.97$.}
\label{fig:sim}
\end{figure}

Figure~\ref{fig:bif_prof}(a) shows that the model admits a SNIPER bifurcation at $\rho\equiv \rho_{\rm SNIPER} \simeq 0.99005$, with a spatially uniform limit cycle $\vP_{\rm osc}(t\in[0,\tau])\equiv(A_{\rm osc},H_{\rm osc},S_{\rm osc},Y_{\rm osc})^{\rm T}$ present for $\rho<\rho_{\rm SNIPER}$ [panel (b)]. Besides the neutral Floquet multiplier $\lambda_0=1$, there are three real multipliers, all smaller than $\mathcal{O}(10^{-5})$, extending all the way to $\rho=0$. The aforementioned spiking is found in the regime $\rho_{\rm SNIPER}<\rho<\rho_{\rm Turing}$~\cite{knobloch2024emergence}, much as in Fig.~\ref{fig:theta}(c), and persists into $\rho \lesssim \rho_{\rm SNIPER}$, suggesting that $\vP_{\rm osc}$ is unstable in this regime, as shown in Fig.~\ref{fig:sim}, cf.~Fig.~\ref{fig:theta}(b). 

We use the numerical solution for $\vP_{\rm osc}(t\in[0,\tau])$, obtained via numerical continuation, to evaluate $\sigma_1(\rho)$ using~\eqref{eq:sig1} and $\widetilde \sigma_1(\rho)$ using~\eqref{eq:sigma1_apprx}: the former predicts onset of the long-wavelength instability at $\rho=\rho_{\rm LW}\simeq 0.984$, as shown by the red line in Fig.~\ref{fig:bif_prof}(a) (right $y$-axis); the latter gives $\rho=\rho_{\rm c}\simeq 0.981$. In Fig.~\ref{fig:bif_prof}(b), we show selected profiles of the limit cycle $\vP_{\rm osc}(t)$ in the vicinity of $\rho_{\rm LW}$. In SM, we compare the results with direct numerical computation of the onset using monodromy matrix eigenvalues and find good agreement with $\rho_{\rm LW}$.

To confirm these predictions, we employ DNS~\cite{num} on a representative domain, $L=60$, in two cases: $\rho_{\rm LW}<\rho<\rho_{\rm SNIPER}$ [Fig.~\ref{fig:sim}(a)] and $\rho < \rho_{\rm LW}$ [Figs.~\ref{fig:sim}(b,c)]. Figure~\ref{fig:sim}(a) shows an abrupt breakup of a spatially uniform oscillation into irregular spiking, while Fig.~\ref{fig:sim}(b) shows that the breakup into spiking persists for lower $\rho$, but is both delayed and develops through a long-wavelength instability that generates an asynchronous oscillon lattice. The spiking appears to be triggered in regions of negative curvature of the oscillating stripe and is thus a consequence of the breakdown of synchrony (see Fig.~S3 in SM). The spiking period is slightly shorter than the oscillation period $\tau$, thereby amplifying the synchrony breakdown, which grows in spatial extent at a well-defined speed \cite{knobloch2024emergence}. The breakdown process imprints a well-defined lattice wavelength, $\nu\approx 5$, on the spiking dynamics, prior to its eventual disruption [Fig.~\ref{fig:sim}(b)] and the emergence of a sea of jumping oscillons~\cite{yang2006jumping,knobloch2021origin,knobloch2024emergence}). We conjecture that the initial lattice spacing correlates with the local inhibition of spiking in the vicinity of each oscillon on a length scale $\zeta\equiv\sqrt{D_{\rm H}/D_{\rm A}}\approx 4.5$. For example, when $\rho=0.987$ {and $D_{\rm H}=0.02$ ($\zeta\approx 6.3$, $D_{\rm H}=0.04$), we find that in a domain with $L=5$ ($L=8$) and Neumann boundary conditions the spikes are attached to the two boundaries (and oscillate out of phase) but when $L=6$ ($L=9$)} the spikes start to detach from the boundaries and the spiking becomes asynchronous (not shown), indicating that in larger domains the oscillon asynchrony ultimately disrupts the oscillon lattice as the spacing between adjacent spikes changes over time. This is a slow process since the spikes interact only via their tails. Since the spatial instability is far from marginal the perturbations grow fast, resembling the situation in Fig.~\ref{fig:sim}(a). Very close to the SNIPER, however, the spikes emerge periodically with a wavelength comparable to the fastest growing mode at the fold, $2\pi/k_{\rm max}\approx 2.7$ [Fig.~\ref{fig:sim}(a)]. This is consistent with the observation for the theta model.

{Figures~\ref{fig:sim}(b,c) demonstrate that in large domains the long wave instability is subcritical:} large enough perturbations lead to the emergence of jumping oscillons for $\rho_{\rm s}\approx0.97<\rho<\rho_{\rm LW}$ [Fig.~\ref{fig:sim}(b)] via curvature-driven instability or to spatially modulated oscillations [Fig.~\ref{fig:sim}(c)]. To determine the onset of spiking, $\rho=\rho_{\rm s}$, we used DNS with $L=60$, starting from jumping oscillons above $\rho_{\rm LW}$ and decreasing $\rho$ adiabatically to $\rho\lesssim \rho_{\rm s}=0.965$. This resulted in a transition from jumping oscillons to modulated oscillations. We then used the modulated oscillations as initial conditions for $\rho_{\rm s}<\rho<\rho_{\rm LW}$ and found the time, $t_{\rm s}$, at which these break up into spikes. In Fig.~\ref{fig:sim}(d), we show that $t_{\rm s}\sim 1/\sqrt{|\rho-\rho_{\rm s}|}$, a scaling that is typical {of the vicinity of a finite wavenumber instability onset such as the Turing instability~\cite{knobloch2024emergence}}. Overall, the results indicate that the subcritical instability at $\rho_{\rm LW}$ is the result of front curvature that persists from $\rho=0$ to $\rho_{\rm s}$, where the fronts become unstable and break down into spatiotemporal periodic spiking in the form of jumping oscillons. 

\textit{Discussion.--} Understanding the mechanisms generating spatiotemporal patterns, especially those that are described by deterministic non-gradient PDEs~\cite{golubitsky1999pattern}, is of paramount importance in applications at all scales~\cite{whitesides2002self,meron2015book,cross2009pattern,pismen2023patterns}. In the case of large amplitude spatially uniform relaxation oscillations emerging near global bifurcations~\cite{izhikevich2007dynamical,strogatz2018nonlinear}, the instability is of long-wavelength type, but is typically determined in a cumbersome fashion~\cite{fairgrieve1991ok,pena2007two,gallay2011diffusive}. As a result, investigation of nonlinear PDE models exhibiting a SNIPER bifurcation, a bifurcation associated with the transition from large amplitude relaxation oscillations to (type-I) excitability~\cite{gaspar1986bifurcation,bar1987relevance,dubbeldam1999excitability,ermentrout2019recent,schelte2020dispersive,moreno2022bifurcation}, is considered a major challenge in pattern formation theory~\cite{kuramoto1984chemical,tuckerman2020computational}. 

We presented a robust route to spatially nonuniform relaxation dynamics in non-gradient systems and demonstrated that a nearby Turing instability of uniform steady states may be imprinted on unstable uniform oscillations near a SNIPER bifurcation and so be responsible for the presence of nonuniform oscillations in a region where uniform oscillations are absent. We have exhibited two examples of multi-variable systems, one with chimera-like states and one with spatiotemporal spiking, to emphasize that the realization of such oscillations may be model-dependent. {We believe that these results are relevant not only to understanding oscillations in type-I excitable systems but in view of the simplicity of the system~\eqref{eq:theta} to advancing the study of chimera states in spatially extended systems~\cite{panaggio2015chimera,yao2016chimera,omel2018mathematics,majhi2019chimera,parastesh2021chimeras,ferre2023critical}. In particular, we point to the potential application of our results to understanding epileptic seizures~\cite{andrzejak2016all,lainscsek2019cortical,azad2025leveraging,bartels2025event,andrzejak2026seizure}}. 
%
\newline \newline
\noindent{\bf Acknowledgments.} We thank Svetlana Gurevich (M{\"u}nster) for helpful discussions and anonymous referees for their comments. This work was supported by grant no. 2022072 from the United States--Israel Binational Science Foundation (BSF), Jerusalem, Israel.

\bibliography{branching-bib.bib}

\clearpage
\onecolumngrid
\setcounter{page}{1}
\setcounter{figure}{0}
\setcounter{equation}{0}
\renewcommand{\thesection}{S\arabic{section}} 
\renewcommand{\thesubsection}{S\arabic{subsection}}
\renewcommand{\thefigure}{S\arabic{figure}}
\renewcommand{\thepage}{S\arabic{page}}
\renewcommand{\theequation}{S\arabic{equation}}
\patchcmd{\subsection}{\centering}{\raggedright}{}{}

\section*{Supplementary Material}\label{sec:SM}
\subsection{Spatially extended scalar model for a SNIPER}

We consider the scalar spatial partial differential equation (PDE):
\begin{equation}
	\frac{\partial \theta}{\partial t} = f(\theta) - a\frac{\partial^2 \theta}{\partial x^2} - b\frac{\partial^4 \theta}{\partial x^4} =1-\cos \theta + \mu\bra{1+\cos \theta } - a\frac{\partial^2 \theta}{\partial x^2} - b\frac{\partial^4 \theta}{\partial x^4},
	\label{eq:pde}
\end{equation}
where $\theta$ is a phase-like variable defined mod $2\pi$ and $a,b>0$. Eq.~\ref{eq:pde} is a gradient (relaxational) flow system and may be written as a 
gradient descent,
\begin{equation}
	\theta_t = -\frac{\delta F}{\delta \theta}, \qquad {\rm where} \qquad
	F[\theta,\mu] = -\int \left[\theta - \sin \theta + \mu\bigl(\theta+\sin \theta \bigr) + \frac{a}{2}\theta_x^2 - \frac{b}{2}\theta_{xx}^2\right] {\rm d}x.
	\label{eq:functional}
\end{equation}
Consequently $F[\theta,\mu]$ is a Lyapunov functional for the dynamics:
\begin{equation}
	\frac{{\rm d}F}{{\rm d}t} = \int \frac{\delta F}{\delta \theta}\,\theta_t\,{\rm d}x
	= -\int \theta_t^2\,{\rm d}x \le 0,
	\label{eq:lyapunov}
\end{equation}
with equality only at steady states. If $F(\theta,\mu)$ is bounded from below this result implies that trajectories relax monotonically toward critical points (minimizers or saddles) of $F(\theta,\mu)$, placing the model in the same structural class as the Swift--Hohenberg equation and distinct from dissipative-but-non-gradient systems such as the Kuramoto--Sivashinsky equation.
However,~\eqref{eq:functional} is not bounded from below as $\theta$ increases, allowing \eqref{eq:pde} to continue reducing its free energy $F$ by increasing $\theta$. With periodic boundary conditions (mod $2\pi$) this type of solution becomes a time-periodic solution -- despite the gradient structure of Eq.~\eqref{eq:pde}.
\begin{figure}[tp!]
	\centering
	(a){\includegraphics[width=0.98\textwidth]{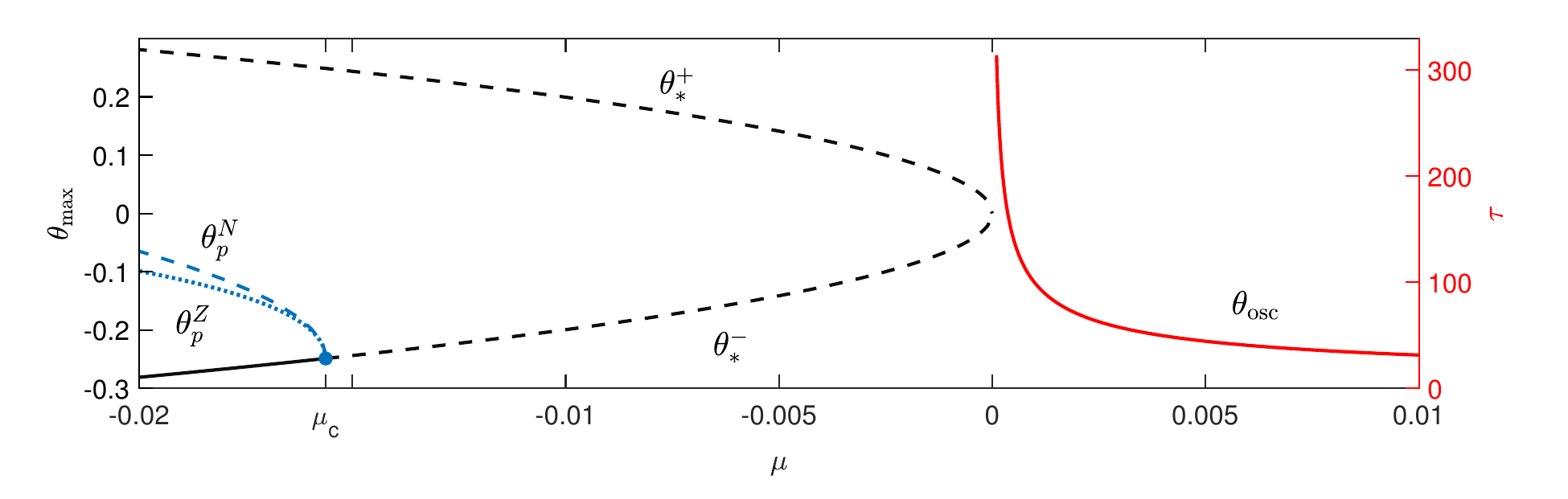}}
	(b){\includegraphics[width=0.95\textwidth]{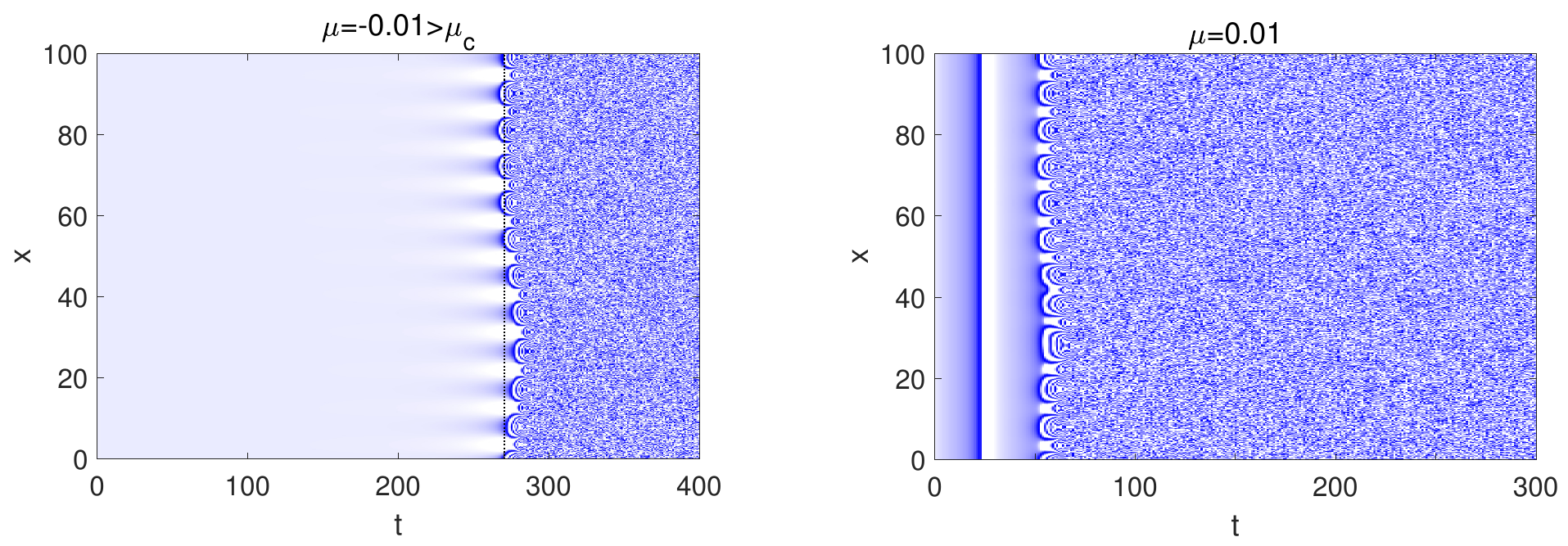}}
	(c){\includegraphics[width=0.95\textwidth]{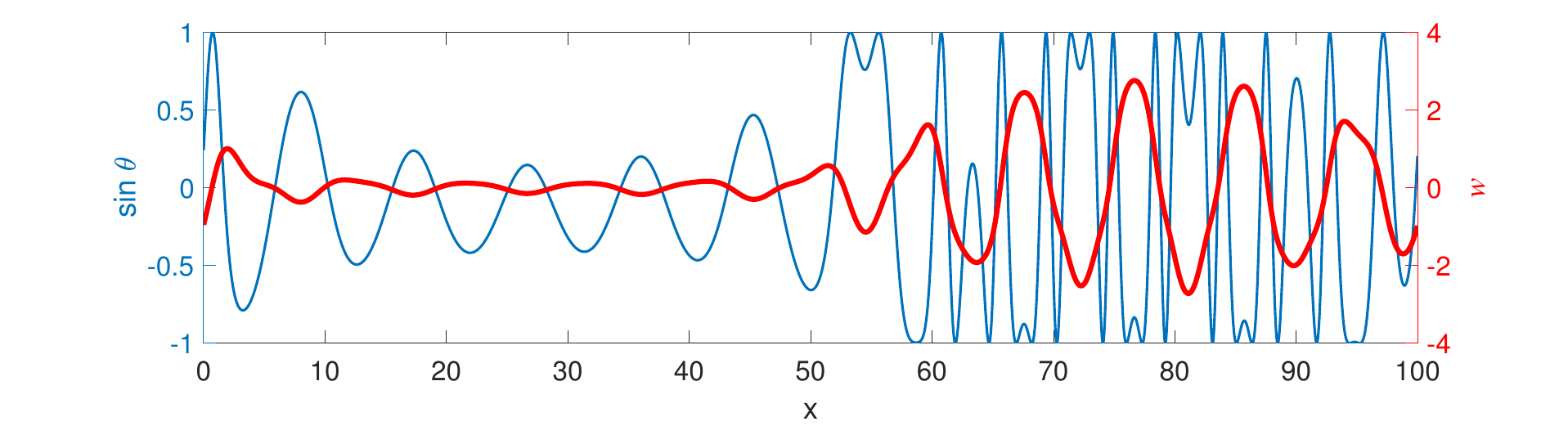}}
	\caption{(a) Bifurcation diagram showing the branches $\theta^\pm_*$ of steady uniform solutions together with the temporal period $\tau=\pi/\sqrt{\mu}$ of uniform oscillations $\theta_{\rm osc}$ (red, associated with the right $y$-axis). Additionally, we plot the steady spatially periodic states ($\theta^{N,Z}_{p}$, where superscripts $N$ and $Z$ refer to branches obtained via numerical continuation and Eq.~\eqref{eq:amp_max}, respectively) that emerge from $\theta_*^-$ at $\mu_{c}=-1/64$ with $k_{c}=1/\sqrt{2}$, for $a=b=1$. Here $\mu_{\rm SNIPER}=0$ corresponds to the SNIPER bifurcation and solid/dashed lines indicate linearly stable/unstable equilibria; $\theta_{\rm osc}$ are linearly unstable for $\mu>\mu_{\rm SNIPER}=0$. (b) DNS of the scalar model \eqref{eq:pde} with periodic boundary conditions on a domain of length $L=100$, showing a space-time plot of $\sin \theta$ for two values of $\mu$. The initial conditions are random $\mathcal{O}(10^{-5})$ fluctuations about either $\theta_*^-\simeq -0.1993$. Both cases initially manifest the wave number $k_c$ but this wave number is erased with increasing time. (c) Profiles of $\sin\theta(x)$ and $w(x)$ for $\mu=-0.01$ at $t=270$ [dotted cross section in (b)], according to~\eqref{eq:scalar_2var}}.
	\label{fig:scalar}
\end{figure}

These time-periodic states are created via the annihilation of a pair of equilibria, i.e., via a SNIPER bifurcation. This bifurcation takes place at $\mu=\mu_{\rm SNIPER}=0$, corresponding to a fold connecting a pair of spatially homogeneous steady states $\theta(x,t)=\theta_*^\pm$ that coexist for $\mu\leq 0$, with $\theta_*^{-}\equiv -\arccos \dfrac{1+\mu}{1-\mu}$ stable and $\theta_*^{+}\equiv \arccos \dfrac{1+\mu}{1-\mu}$ unstable. For $\mu>0$, no uniform steady states exist, and $\theta$ increases monotonically with period $\tau = \pi/\sqrt{\mu}$, $0<\mu\ll1$. The resulting spatially uniform oscillations go through a long wavelength instability with Floquet exponent
\begin{equation}
	\sigma(k) = ak^2-bk^4,
\end{equation}
while for $\mu<0$ perturbations of $\theta_*^-$ obey the dispersion relation
\begin{equation}
	\sigma(k) = -2\sqrt{-\mu} + ak^2 - bk^4,
	\label{eq:dispersion}
\end{equation}
indicating that $\theta_*^-$ loses stability in a steady state bifurcation at $\mu=\mu_{\rm c} \equiv -\dfrac{a^4}{64\,b^2}$ with nonzero critical wave number $k=k_c \equiv  \sqrt{\dfrac{a}{2b}}$. 
To determine the direction of branching of the resulting steady spatially periodic solutions, we employ a standard multiple time scale analysis~\cite{ch93}. We define a small parameter $|\epsilon|\ll1$ such that
$\mu = \mu_{\rm c} + \epsilon^2\mu_2$, a slow time $T=\epsilon^2 t$, and expand $\theta(x,t)$:
\begin{equation}
	\theta(x,t) = \theta_*^- + \epsilon \theta_1(x,T) + \epsilon^2 \theta_2(x,T) + \epsilon^3 \theta_3(x,T) + O\bra{\epsilon^4},
	\label{eq:expansion}
\end{equation}
where
\[
\theta_*^-(\mu)=\theta_*^-(\mu_{\rm c})+\epsilon^2\mu_2
\left.\frac{{\rm d}\theta_*^-}{{\rm d}\mu}\right|_{\mu_{\rm c}}+O\bra{\epsilon^4} \qquad
{\rm and} \qquad
\theta_1(x,T) = Z(T)e^{ik_cx} + \bar Z(T)e^{-ik_cx}.
\] 
Finally, the linear operator applied through the different orders is:
\begin{equation}
	\mathcal{L}_c = f'(\theta_*^-) - a\frac{\partial^2}{\partial x^2} - b\frac{\partial^4}{\partial x^4}
	= -\frac{a^2}{4b} - a\frac{\partial^2}{\partial x^2} - b\frac{\partial^4}{\partial x^4}.
	\label{eq:Lc}
\end{equation}
Omitting all algebraic steps at different orders in $\eps$, we end up with a real-valued Stuart-Landau amplitude equation
\begin{equation}
	\frac{\partial Z}{\partial T} = \frac{8b}{a^2}\mu_2\, Z + \frac{9a^4+304b^2\left(1+\mu_c\right)^2}
	{72a^2b}\ABS{Z}^2Z.
	\label{eq:landau}
\end{equation}
Thus, for $a,b>0$, the coefficient of the cubic term is positive and therefore the instability is unconditionally subcritical, i.e., the branch bifurcates in the direction $\mu_2<0$ or equivalently $\mu<\mu_c$, with amplitude
\begin{equation}
	\theta^Z_{p,\max}(\mu)=\theta_*^-(\mu)+
	\frac{48b}
	{\sqrt{9a^4+304b^2\left(1+\mu_c\right)^2}}
	\sqrt{\mu_c-\mu}.
	\label{eq:amp_max}
\end{equation}
The resulting bifurcation diagram is thus similar to that in Fig. 1(a) of the main text [Fig.~\ref{fig:scalar}(a)]. However, in contrast to Figs. 1(b,c) no chimera states are present.

In Fig.~\ref{fig:scalar}(b), we employ DNS for representative parameters $a=b=1$ and show, using space-time plots, the evolution into uniformly distributed chaotic dynamics in the unstable regimes across the SNIPER bifurcation. For comparison with Fig.~1 of the main text, we write~\eqref{eq:pde} as a two-variable system,
\begin{equation}
	\frac{\partial \theta}{\partial t} = f(\theta) - a \frac{\partial^2 \theta}{\partial x^2} - b \frac{\partial^2 w}{\partial x^2}, \qquad \qquad
	w =  \frac{\partial^2 \theta}{\partial x^2},
	\label{eq:scalar_2var}
\end{equation}
and present profiles of both $\theta(x)$ and $w(x)$ at $t=270$ [Fig.~\ref{fig:scalar}(c)]. The figure shows that in contrast to Fig.~1(d) of the main text $w(x)$ has simple maxima.

We have integrated~\eqref{eq:pde} for other parameters also, but have found no persistent nonuniform spatiotemporal oscillations.

\subsection{Meinhardt's model equations}
The explicit activator-inhibitor-substrate model reads~\cite{meinhardt1976morphogenesis}:
\begin{subequations}\label{eq:AImodel}
	\begin{eqnarray}
		\frac{\partial A}{\partial t}&=&c\dfrac{SA^2}{H}-\mu A+\rho_{\text{A}} Y+D_{\text{A}} \dparf{A}{x}, \\
		\frac{\partial H}{\partial t}&=&cSA^2-\nu H+\rho_{\text{H}} Y+D_{\text{H}} \dparf{H}{x}, \\
		\frac{\partial S}{\partial t}&=&c_0-\gamma S-\varepsilon Y S+D_{\text{S}} \dparf{S}{x}, \\
		\frac{\partial Y}{\partial t}&=&d A-eY+\dfrac{Y^2}{1+fY^2}+D_{\text{Y}} \dparf{Y}{x}.
	\end{eqnarray}
\end{subequations}
We follow earlier studies (\cite{knobloch2024emergence} and references therein), and adopt the parameters $c=0.002$, $\mu=0.16$, $\rho_{\text{A}}=0.005$, $\nu=0.04$, $c_0=0.02$, $\gamma=0.02$, $\varepsilon=0.1$, $d=0.008$, $e=0.1$, $f=10$, $D_{\text{A}}=0.001$, $D_{\text{H}}=0.02$, $D_{\text{S}}=0.01$, $D_{\text{Y}}=10^{-7}$, and vary $\rho\equiv\rho_{\text{H}}10^{5}$ as a control/bifurcation parameter.
\begin{figure}[tp]
	\centering
	(a){\includegraphics[width=0.47\columnwidth]{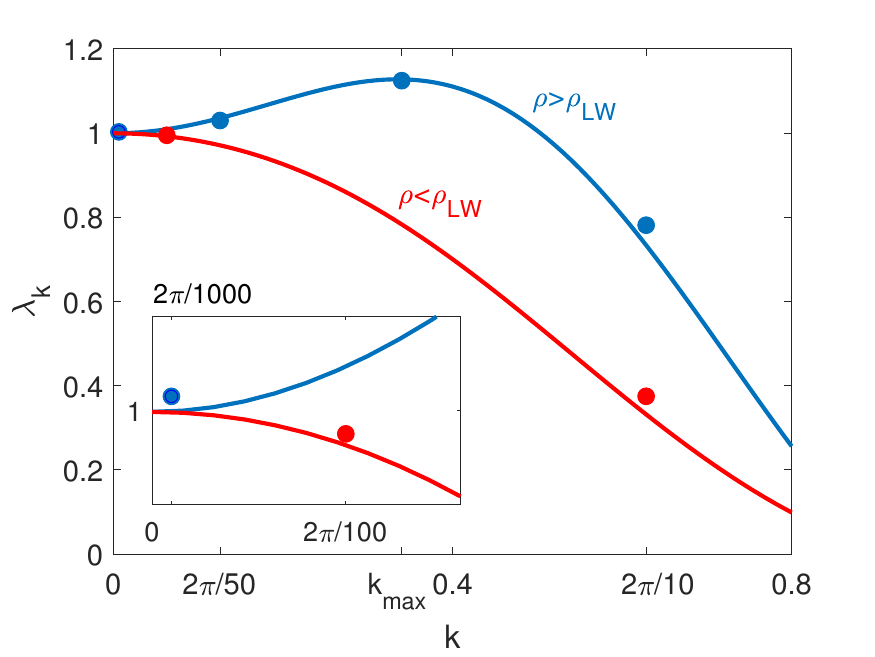}}
	(b){\includegraphics[width=0.47\columnwidth]{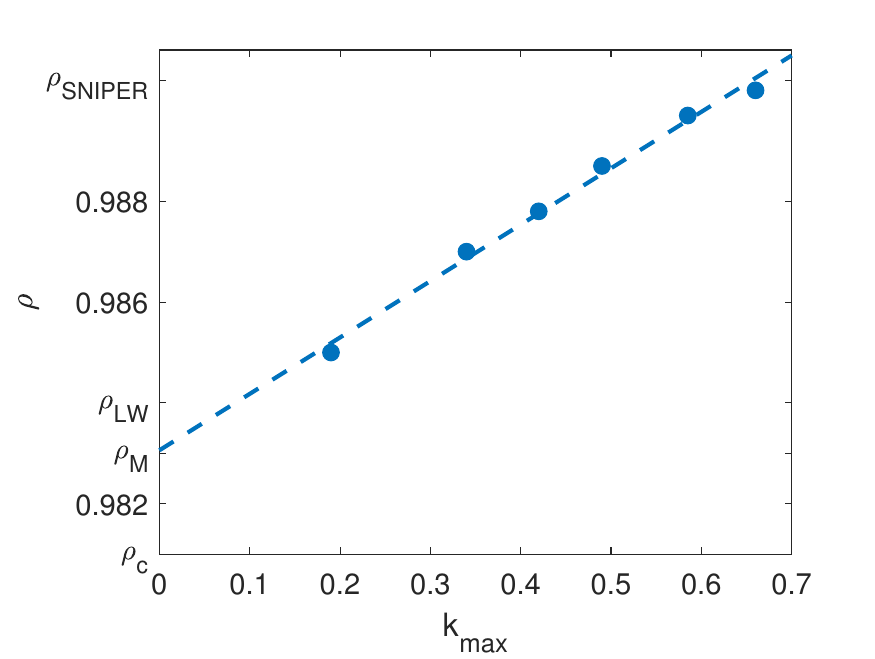}}
	\caption{(a) Dispersion relation for the Meinhardt model \eqref{eq:AImodel} showing the leading Floquet multiplier $\lambda_{k}$ as a function of the wave number $k$ computed for $\rho=0.98<\rho_{\rm LW}\approx 0.984$ (red line) and $\rho=0.987>\rho_{\rm LW}$ (blue line) via eigenvalue analysis of the monodromy matrix according to~\eqref{eq:mon_mat} with (2), whereas $\rho_{\rm LW}$ is computed via (4) with~\eqref{eq:L0},~\eqref{eq:p0}, and~\eqref{eq:hat_p0}. The dots correspond to Floquet multipliers with spatial period $L$, computed numerically using~\eqref{eq:mon_mat} with a second-order finite difference Laplacian and periodic boundary conditions at selected domain lengths $L=10,18.48,50,100,1000$, where $L=18.48=2\pi/k_{\max}$. (b) Plot of $k_{\rm max}$, the wave number corresponding to maximum growth rate, as a function of $\rho$; the linear fit (dashed line) suggests that $\rho(0)\equiv\rho_{\rm M}\approx 0.983$, in good agreement with $\rho_{\rm LW}\approx 0.984$.}
	\label{fig:LW_inst}
\end{figure}

\subsection{Long-wavelength dispersion relation computed via monodromy matrix}
We also compare the instability onset obtained through (4) with a direct monodromy matrix eigenvalue analysis, using two forms of (2): 
\begin{equation}\label{eq:mon_mat}
	\partial_t{\Phi}(t,x)=\mathcal{L}(t,\partial_x)\Phi(t,x),
\end{equation}
where $\Phi$ is the fundamental matrix solution, with $\Phi(0,k)=\mathbb{I}$, for the linearized equation about the uniform limit cycle $\vP_{\rm osc}$. In this setting, the eigenvalues of the spatially extended monodromy matrix $\mathcal{M}\equiv \Phi(\tau,k)$, where $k$ is the wave number, 
correspond to the Floquet multipliers~\cite{fairgrieve1991ok,hartman2002ordinary,pena2007two,juma2026using}. In case (\textit{i}), we use (2) with~\eqref{eq:L0} and iterate $k$ to obtain the dispersion relation and hence the leading Floquet multiplier $\lambda_{k}$. In case (\textit{ii}), we use $\mL{}{}=\mL{0}{}+\mathcal{D}\partial^2_x$, where the spatial derivative is evaluated on periodic domains of different lengths, resulting in a large matrix of four square blocks, where each block depends on the number of grid points. Our results are summarized in Fig.~\ref{fig:LW_inst}, with $\rho(0)\equiv\rho_{\rm M}\approx0.983$ corresponding to the computed onset. The figure shows that both sets of results not only agree with the onset $\rho_{\rm LW}$ obtained from Eq. 3 but also confirm that the instability is indeed of the long-wavelength type.

For the system~\eqref{eq:AImodel} we have the explicit expressions
\begin{equation}\label{eq:L0}
	\mL{0}{}\equiv\begin{pmatrix}
		\mL{1}{1} & \mL{1}{2} & \mL{1}{3} & \mL{1}{4} \\
		\mL{2}{1} & \mL{2}{2} & \mL{2}{3} & \mL{2}{4}  \\
		0 & 0 & \mL{3}{3} & \mL{3}{4}  \\
		\mL{4}{1} & 0 & 0 & \mL{4}{4}  
	\end{pmatrix},
\end{equation}
where $\mL{1}{1}={2 c A_{\rm osc} S_{\rm osc}}/{H_{\rm osc}} - \mu$, $\mL{1}{2}= -{c A^2_{\rm osc} S_{\rm osc}}/{H^2_{\rm osc}}$, $\mL{1}{3}={c A^2_{\rm osc}}/{H_{\rm osc}}$, $\mL{1}{4}=\rho_{\rm A}$, $\mL{2}{1}=2 c A_{\rm osc} S_{\rm osc}$, $\mL{2}{2}= -\nu$, $\mL{2}{3}= c A_{\rm osc}^2$, $\mL{2}{4}= \rho_{\rm H}$, $\mL{3}{3}= -\gamma - \varepsilon Y_{\rm osc}$, $\mL{3}{4}= -\varepsilon S_{\rm osc}$, $\mL{4}{1}=d$, $\mL{4}{4}= -e+{2 Y_{\rm osc}}/{\bra{1 + f Y^2_{\rm osc}}^2}$. Note that $\mL{0}{}$ has almost a triangular block structure and kernel vectors with three components enslaved
\begin{equation}\label{eq:p0}
	{\bf v}_0\bra{t}=\begin{pmatrix}
		v_1 \\
		v_2 \\
		v_3 \\
		v_4     
	\end{pmatrix}=\begin{pmatrix}
		-{\mL{4}{4}}/{\mL{4}{1}} \\
		-\bra{\mL{1}{1}v_1+\mL{1}{3}v_3+\mL{1}{4}}/{\mL{1}{2}} \\
		-{\mL{3}{4}}/{\mL{3}{3}} \\
		1 
	\end{pmatrix},
\end{equation}
and since $\mL{0}{}^{\dagger}=\mL{0}{}^{\rm T}$
\begin{equation}\label{eq:hat_p0}
	{\widehat {\bf v}}_0\bra{t}=\begin{pmatrix}
		\widehat v_1 \\
		\widehat v_2 \\
		\widehat v_3 \\
		\widehat v_4 
	\end{pmatrix}=\begin{pmatrix}
		1 \\
		-{\mL{1}{2}}/{\mL{2}{2}} \\
		-\bra{\mL{1}{3}+\mL{2}{3} \widehat v_2}/{\mL{3}{3}} \\
		-\bra{\mL{1}{1}+\mL{2}{1} \widehat v_2}/{\mL{4}{1}} 
	\end{pmatrix}.
\end{equation}

\clearpage

\subsection{Breakup of a curved relaxation front}

\begin{figure}[ht!]
	\centering
	{\includegraphics[width=0.85\textwidth]{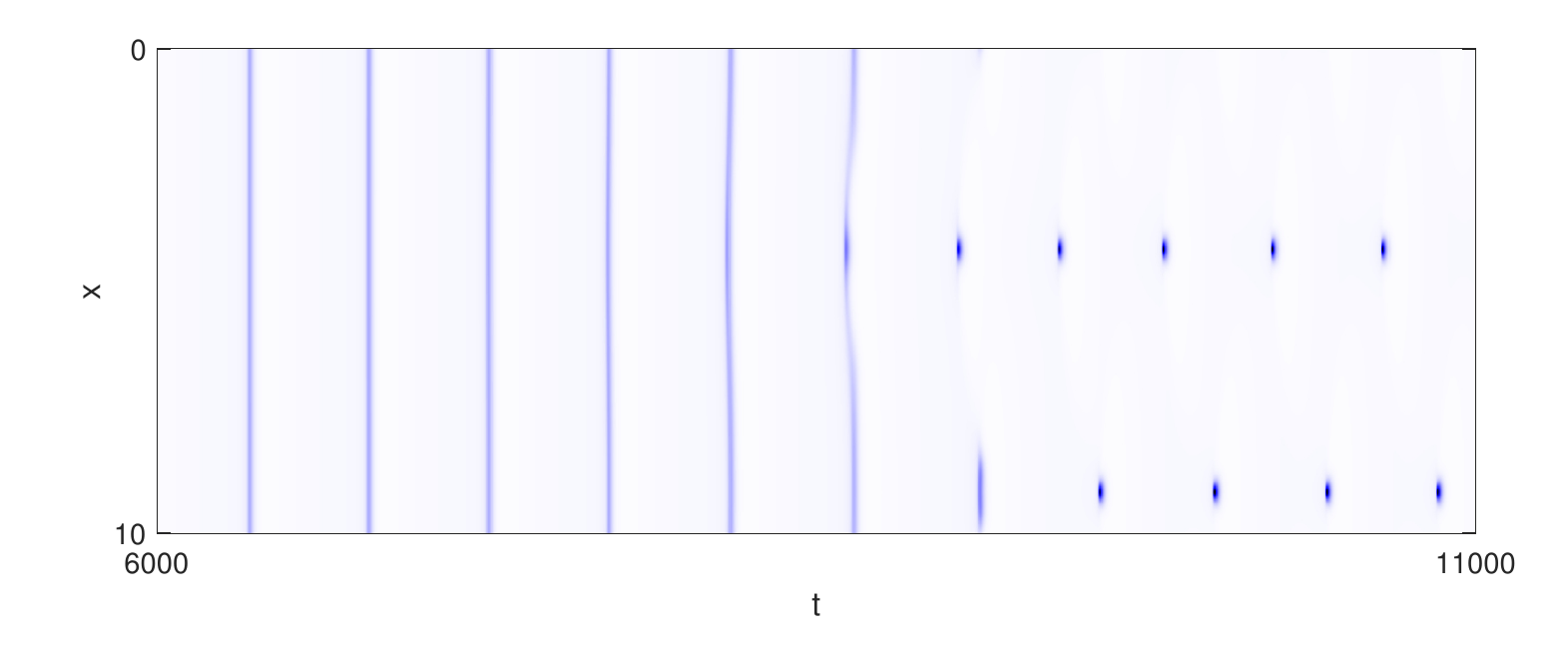}}
	\caption{DNS of the Meinhardt model \eqref{eq:AImodel} with periodic boundary conditions on a domain of length $L=10$, showing a space-time plot of $A(x,t)$ at $\rho_{\rm LW}<\rho=0.987<\rho_{\rm SNIPER}$. The initial conditions are random $\mathcal{O}(10^{-5})$ perturbations of the uniform oscillation amplitude in the fields $A,S,Y$; the field $H$ is not perturbed.}
	\label{fig:sim_SM}
\end{figure}

To further demonstrate the breakup into spikes in regions of slightly negative curvature of an oscillating front, we employ DNS of the Meinhardt model \eqref{eq:AImodel} on a representative domain $L=10$ at $\rho_{\rm LW}<\rho=0.987<\rho_{\rm SNIPER}$, as shown in Fig.~\ref{fig:sim_SM}.

\end{document}